\documentclass[reprint,aps,prl,twocolumn,showpacs,superscriptaddress]{revtex4-1}
\usepackage[utf8]{inputenc}
\usepackage[english]{babel}
\usepackage{amsmath}
\usepackage{amsfonts}
\usepackage{amssymb}
\usepackage{bm}
\usepackage{siunitx}	
\usepackage{graphicx}
\usepackage{natbib}
\newcommand{\Chem}[1]{$\mathrm{#1}$} % Chemical formulas

\begin{document}

% off diagonal K from alloy disorder
% VCA band structure the same as averaged. Justifies use of effective medium
% define K from ifc
% show Marco sandips solution limits papers
\title{First-principles quantitative prediction of the lattice thermal conductivity in random semiconductor alloys: the role of force-constant disorder.} % Force line breaks with \\

\author{Marco Arrigoni}
 \email{marco.arrigoni@tuwien.ac.at}
\author{Jesús Carrete}
\affiliation{%
Institute of Materials Chemistry, TU Wien, A-1060 Vienna, Austria
}%
\author{Natalio Mingo}
\affiliation{%
CEA, LITEN, 17 rue des Martyrs, F-38054 Grenoble, France
}
\author{Georg K. H. Madsen}
\affiliation{%
Institute of Materials Chemistry, TU Wien, A-1060 Vienna, Austria
}%

\date{\today}% It is always \today, today,

\begin{abstract}
The standard theoretical understanding of the lattice thermal conductivity, $\kappa_{\ell}$, of semiconductor alloys assumes that mass disorder is the most important source of phonon scattering. In contrast, we show that the hitherto neglected contribution of force-constant (IFC) disorder is essential to accurately predict the $\kappa_{\ell}$ of those polar compounds characterized by a complex atomic-scale structure.
We have developed an \emph{ab initio} method based on special quasirandom structures and Green's functions, and including the role of IFC disorder, and applied it in order to calculate the $\kappa_{\ell}$ of \Chem{In_{1-x}Ga_xAs} and \Chem{Si_{1-x}Ge_x} alloys. We show that, while for \Chem{Si_{1-x}Ge_x}, phonon-alloy scattering is dominated by mass disorder, for \Chem{In_{1-x}Ga_xAs}, the inclusion of IFC disorder is fundamental to accurately reproduce the experimentally observed $\kappa_{\ell}$. As the presence of a complex atomic-scale structure is common to most III-V and II-VI random semiconductor alloys,  we expect our method to be suitable for a wide class of materials.
\end{abstract}

\maketitle

Random semiconductor alloys are receiving a considerable amount of attention due to their central role in a wide range of technologies, such as photonics \cite{Chen2011,Yao2012}, electronics and optoelectronics \cite{Mokka2009,Park2010,Itzler2011}. It has been observed that important physical quantities, such as the cell parameters and electronic band gap,  can be made to vary continuously between the limiting values of the parent compounds \cite{Schnohr2015}, making the possibility of tuning the alloy properties through the component concentrations of particular interest.

The determination of the thermal conductivity, $\kappa$, is an essential part for the design of all power-dissipating devices, such as lasers, diodes and transistors, based on these alloys.
The ability to accurately calculate $\kappa$ from first principles is, therefore, particularly attractive, as it can significantly help the discovery and design of materials with desirable thermal properties.
 In the last decade, the combination of density functional theory (DFT) with the Boltzman transport equation (BTE) has demonstrated to be a reliable method for accurately determine $\kappa$ of many semiconductor and insulator materials. In these systems, phonons are the main heat carriers and the principal contribution to $\kappa$ is the lattice thermal conductivity, $\kappa_{\ell}$, which can be obtained from the \emph{ab initio} computation of  interatomic force constants (IFCs) \cite{Mingo_chapt,ShengBTE_2014}. 
 While methods based on this approach are nowadays well established for calculating the  $\kappa_{\ell}$ of single crystals \cite{Broido2007,Ward2009,Lindsay2012,Ma2014,Vermeersch2015} and are showing promising results for crystals with point-like and extended defects \cite{Katcho2014,Katre2016,Katre2017,Wang2017}, a method able to correctly describe $\kappa_{\ell}$ of general random semiconductor  alloys is still missing. Hitherto the most commonly employed approach  is based on the virtual crystal approximation (VCA). It consists in describing the random alloy by an effective medium whose properties (lattice constants, IFCs, masses, etc.) are given by the concentration  average of the equivalent properties in the parent compounds. The thermal conductivity is then calculated from the averaged IFCs and the effect of the alloy disorder on phonon transport is taken into account by  introducing  a mass perturbation in an approach analogous to the one employed by Tamura in the study of phonon scattering due to isotopic disorder \cite{Tamura1983}. This method has been successfully employed in calculating the  $\kappa_{\ell}$ of materials such as \Chem{Si_{1-x}Ge_x} \cite{Jivtesh2011} and \Chem{Mg_2Si_{1-x}Sn_x} \cite{Wu2012}. However, it is not adequate to describe III-V and II-VI random semiconductor alloys, as one may see from the comparison between the \emph{ab initio} data for \Chem{In_{1-x}Ga_xAs} \cite{Bjorn2016} and the corresponding experimental values \cite{Hockings1966}. 
This lack of success can be linked to the primary hypothesis of the VCA, namely the representation of an alloy through a non-structural effective medium. In this medium, alloy atoms are placed in the exact same environment as they have in the parent compounds and bond distances are simply assumed to depend linearly on the alloy concentration.  In contrast, experimental observations have found that the atomic-scale structure of semiconductor alloys is actually characterized by large fluctuations from the average-medium structure of the VCA \cite{Schnohr2015}. The presence of this structural disorder at the atomic scale will greatly affect the material's IFCs in a way that cannot be modeled by simply averaging over the pure-compound IFCs.

In this contribution, we take  cubic \Chem{In_{1-x}Ga_xAs}, which presents many features common to general  III-V semiconductor alloys \cite{Schnohr2015},  as an example.  Extended X-ray-absorption fine-structure (EXAFS) measurements clearly show that the \Chem{Ga-As} and \Chem{In-As} nearest-neighbor (NN) distances in \Chem{In_{1-x}Ga_xAs} are mostly unaffected by the alloy concentration and assume the same values as in the parent compounds \cite{Mikkelsen1982}. In addition, the \Chem{As-As} NN distances show a bimodal distribution, while cation-cation distances, which agree better with the VCA, are still distributed over a somewhat broad range  \cite{Mikkelsen1982}. The presence of this structural disorder at the atomic scale is not unique to \Chem{In_{1-x}Ga_xAs}, but is observed in most  III-V and II-VI random semiconductor alloys \cite{Schnohr2015}.  

By directly taking into account the local atomic-scale structural disorder, we show that a big improvement over the VCA can be achieved. This method has allowed us to accurately reproduce the experimentally observed $\kappa_{\ell}$ of \Chem{In_{1-x}Ga_xAs} random alloys at room temperature, with a relative error of around 10 \%, an accuracy comparable to the best predictions achieved for single crystals. On the other hand, we have found that  in \Chem{Si_{1-x}Ge_x} random alloys the IFC disorder does not have a relevant effect on the phonon-alloy elastic scattering rates, which explains why non-structural models such as the VCA are adequate for this compound. The $\kappa_{\ell}$ calculated for \Chem{Si_{1-x}Ge_x} with our approach differs from the one calculated with the VCA by less than 10 \%, indicating that the two approaches yield compatible results in the limit where alloy disorder can be described just through a mass perturbation.

The effect of the alloy atomic-scale structure on the material's IFCs can be readily taken into account from first-principles by employing special quasirandom structures (SQS) \cite{Zunger1990}. SQSs are supercells built in such a way that the correlation functions of a given set  of atomic clusters (pairs, triplets, etc.) match those of a true random alloy as closely as possible. This approach, unlike non-structural models, allows for a direct description of the local atomic structure and for a quantitative evaluation of the effect of structural disorder on the system's IFCs. 
The use of SQSs in the first-principles DFT computation of second-order IFCs has been demonstrated in metallic alloys, where SQS containing as few as 32 or 64 atoms allowed for an accurate reproduction of the phonon band structure  \cite{Wang2011,Chouhan2014}.
We built SQSs representing \Chem{In_{1-x}Ga_xAs} and \Chem{Si_{1-x}Ge_x} random alloys employing the \texttt{mcsqs} module of the Alloy Theoretic Automated Toolkit \cite{vandeWalle2013}. For both materials, the parent compounds have closely related cubic structures (zincblende for \Chem{GaAs} and \Chem{InAs} and diamond for \Chem{Si} and \Chem{Ge}) with a rhombohedral primitive cell containing two atoms. We considered a $4\times4\times4$ supercell expansion of this primitive cell (for a total of 128 atoms per supercell). Furthermore, SQSs with 250 atoms, corresponding to a  $5\times5\times5$ expansion of the primitive cell, were built in order to evaluate the convergence behavior with respect to the supercell size. A 250-atom supercell was also built for the \Chem{In_{0.75}Ga_{0.25}As} alloy, since at this composition, we found that a $4\times4\times4$ supercell is not able to correctly predict the system phonon band structure in agreement with earlier observations for the parent InAs compound \cite{Bjorn2016}.
The SQS cell parameters were obtained from those of the parent compounds following  Vegard's law and were kept fixed during the geometrical optimization of the ionic positions. All first-principles calculations were carried out  in the local density approximation  \cite{Perdew1981} with the projector augmented-wave method \cite{Blochl1994} as implemented in the package VASP \cite{Kresse1999}. Second-order IFCs were calculated employing the finite displacement approach. The \texttt{phonopy} code \cite{phonopy} was used to generate a set of  6 displacements for each atom in the SQS  and to extract the IFCs from the calculated atomic forces. More details on the computational setup and SQS construction can be found in the Supplementary Materials.

Figure \ref{fig:bond_distr} displays the bond length distributions obtained after relaxing the atomic positions in the SQSs representing \Chem{In_{0.3}Ga_{0.7}As} and \Chem{Si_{0.3}Ge_{0.7}} random alloys. As one can see, all the main features in  the atomic-scale structure  observed in the EXAFS experiments can be obtained from the SQS supercells. \Chem{In_{1-x}Ga_xAs} NN distances show two distinct peaks and fluctuations from the structure of the  VCA medium are noticeable even beyond the first coordination shell.  In particular,  the presence of the experimentally observed \cite{Mikkelsen1982} bimodal distribution of the \Chem{As-As} bond lengths is evident, and, even if cation-cation and cation-anion bond distances beyond NNs show a single peak as the VCA would predict, their distribution is broader than in \Chem{Si_{1-x}Ge_x}.
For \Chem{Si_{1-x}Ge_x} we can notice that, except for NN distances, where  \Chem{Si-Si} and  \Chem{Ge-Ge} bond lengths peak at the same value as in the parent compounds, the VCA can reproduce the structure of \Chem{Si_{1-x}Ge_x}.  Specifically, the NN distances possess a small tendency to a bimodal distribution, in agreement with experiments \cite{Hiroshi1992,Aubry1999}, but to a lesser degree than in \Chem{In_{1-x}Ga_xAs}. Bond distances beyond the NN ones are, on the other hand, concentrated around a single peak, independently of the chemical nature of the elements forming the bond.
\begin{figure}[bt]
\centering
\includegraphics[width=0.49\textwidth]{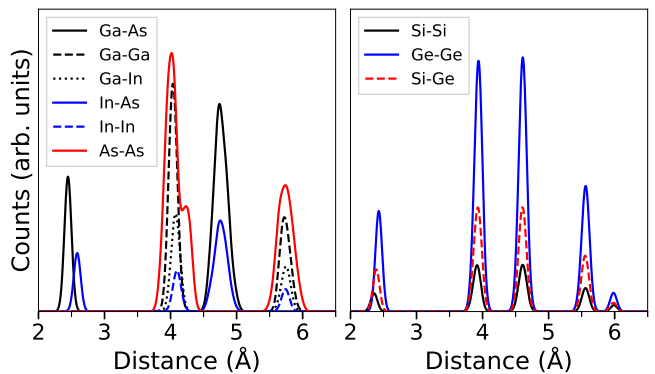} % The . refers to the current directory of the main tex file
\caption{Bond length distribution in \Chem{In_{0.3}Ga_{0.7}As} (left) and \Chem{Si_{0.3}Ge_{0.7}} (right) obtained from the atomic distances calculated with SQSs after applying a Gaussian broadening. }
\label{fig:bond_distr}
\end{figure}

\begin{figure}[bt]
\centering
\includegraphics[width=0.49\textwidth]{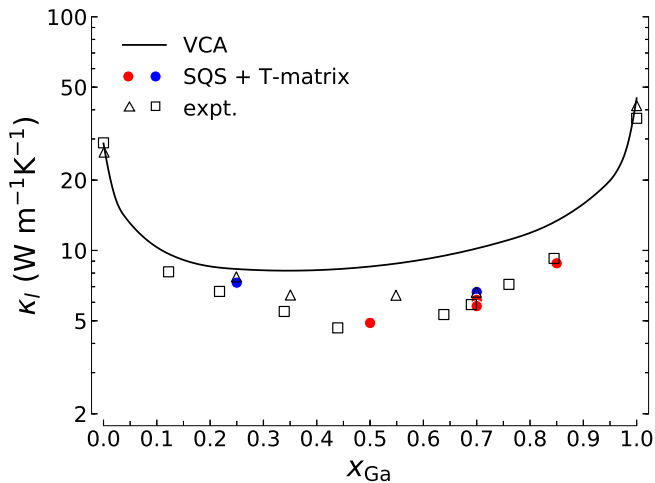} % The . refers to the current directory of the main tex file
\caption{Lattice thermal conductivity, $\kappa_{\ell}$, of \Chem{In_{1-x}Ga_xAs} calculated at \SI{300}{\kelvin} from first-principles with the VCA and with the improved approach based on SQSs. Filled red circles and blue circles represent values obtained using 128-atom and 250-atom SQSs, respectively. The experimental values obtained fom Ref. \onlinecite{Hockings1966} (empty triangles) and Ref. \onlinecite{Abrahams1959} (empty squares) are shown for comparison. }
\label{fig:kappal}
\end{figure}

To emphasize the role of the IFC disorder, we first calculated $\kappa_{\ell}$ for \Chem{In_{1-x}Ga_xAs} using the VCA. Second- and third-order IFCs calculated in a previous study \cite{Vermeersch2015} for \Chem{GaAs} and \Chem{InAs} were taken from the on-line almaBTE database \cite{Alma_database}. 
We limited the consideration of scattering phenomena affecting phonons in bulk semiconductor alloys to three-phonon inelastic scattering and  elastic  scattering between phonons and the alloy disorder. Total scattering rates are given by Matthiessen's rule: $ \tau^{-1} = \tau_{3p}^{-1} + \tau_{dis}^{-1}$, where $\tau_{3p}^{-1}$ and $\tau_{dis}^{-1}$ represent the above-mentioned inelastic and elastic scattering, respectively.  $\tau_{3p}^{-1}$ can be calculated from the VCA second- and third-order IFCs as explained in Refs. \onlinecite{ShengBTE_2014,Broido2007}. The VCA calculates $\tau_{dis}^{-1}$ from Tamura's formula, which treats the alloy disorder on the medium only through the mass perturbation \cite{Tamura1983,Jivtesh2011}. The calculated $\kappa_{\ell}$ in the VCA is reported in Fig.~\ref{fig:kappal} through a bold black line. It is evident that this approximation largely overestimates the measured value of $\kappa_{\ell}$, especially in the region with a \Chem{GaAs} concentration above 40\%.

We then went beyond the mass perturbation picture and directly took into account the IFC disorder using the SQS supercell calculations and  a Green's function perturbative treatment. We consider the alloy disorder as a perturbation affecting an effective medium. The medium is analogous to the one considered in the VCA, but its second-order IFCs are obtained by symmetrizing those calculated from an SQS according to the full space group symmetry of the parent compounds ($\mathrm{F\bar{4}3m}$ for \Chem{In_{1-x}Ga_xAs} and $\mathrm{Fd\bar{3}m}$ for \Chem{Si_{1-x}Ge_x}). As documented in the Supplementary Materials, the resulting harmonic phonon band-structure is very similar to that obtained using the VCA, lending credibility to the effective medium approach. The alloy third-order IFCs, needed for calculating anharmonic three-phonon scattering,  were not obtained from the SQS, but were instead constructed from those of the  parent compounds using the VCA,  avoiding the computational cost involved in calculating the third-order IFCs in systems with a very low symmetry. In addition, the use of the VCA third-order IFCs is justified by their relatively high degree of transferability \cite{Carrete2014}.

The elastic scattering rates, $\tau_{dis}^{-1}$, are obtained considering each atom in the alloy to be responsible for a mass perturbation and, in contrast with the VCA, a perturbation in the second-order IFCs. We considered each chemical element as an independent scatterer, characterized by the perturbation $ \bm{\mathsf{V}}_i = \bm{\mathsf{M}}_i + \bm{\mathsf{K}}_i$ (e.g. in \Chem{In_{1-x}Ga_xAs}, $i$ = Ga, In or As). The mass pertubation is described by the diagonal matrix $\bm{\mathsf{M}}$ and arises from the difference in mass between the actual alloy atom and the VCA atom. The IFC perturbation is described by a matrix $\bm{\mathsf{K}}$ which arises from the disordered alloy structure.
For each scatterer type $i$, the SQS is used to sample the perturbation in the alloy configurational space.
Each perturbation is considered in direct space and its influence  is limited by a cutoff radius, chosen in such a way that the value of  $\tau_{dis}^{-1}$ does not noticeably change if the cutoff is further increased. For \Chem{In_{1-x}Ga_xAs}, we used a cutoff of $\approx$ \SI{6.5}{\angstrom}, corresponding to a perturbation affecting up to the fifth nearest neighbors of the perturbing atom.   $\bm{\mathsf{K}}_i$ is symmetrized according to the  site symmetry that the scatterer possesses in the parent compound. In this way the perturbation is compatible with the effective medium  symmetry and is  localized around its scatterer. Clearly a SQS of infinite size would allow for a complete sampling. In practice the sampling is limited by the system size and can be improved considering different SQS for a given alloy concentration or using larger ones. For instance, our 128-atom SQSs provide  $64\times x_i$ samples for each scatterer.

Once the sampling is done, the averaged perturbation for scatterer $i$, $\langle  \bm{\mathsf{V}}_i \rangle$, is built averaging over the $64\times x_i$ perturbations  $ \bm{\mathsf{V}}_i $ obtained for scatterers of type $i$. 
We found the perturbation averaging to be fundamental in order to reproduce the experimental $\kappa_{\ell}$ of \Chem{In_{1-x}Ga_xAs}, as isolated $\bm{\mathsf{V}}_i$ tend to give a too low $\kappa_{\ell}$. In addition, approximating $\langle  \bm{\mathsf{V}}_i \rangle$ by the perturbation generated by a single impurity of type $i$ in the  parent compound not containing any atom of type $i$ (e.g. \Chem{Ga} in \Chem{InAs} or \Chem{As} in \Chem{GaAs}) also underestimates the value of $\kappa_{\ell}$. This is not surprising, as the atomic-scale structure of a compound containing a single impurity is a very unlikely representative for the complex atomic configurations observed in random semiconductor pseudobinary alloys.

The T-matrices of each scatterer, $\bm{\mathsf{T}}_i$, are then calculated from the averaged perturbations, $\langle  \bm{\mathsf{V}}_i \rangle$, via a direct solution of the Dyson equation: 
\begin{align}
\label{eq:t-matrix}
\begin{aligned}
 \bm{\mathsf{T}}_i %&= \langle \bm{\mathsf{V}}_i\rangle +  \langle \bm{\mathsf{V}}_i \rangle  \bm{\mathsf{g}} \langle \bm{\mathsf{V}}_i \rangle + \cdots \\
&= \left( \bm{\mathsf{I}} - \langle \bm{\mathsf{V}}_i \rangle \bm{\mathsf{g}} \right)^{-1} \langle \bm{\mathsf{V}}_i \rangle,
\end{aligned}
\end{align}
which allows the consideration of scattering processes to all orders. Here $ \bm{\mathsf{I}} $ is the identity matrix and $\bm{\mathsf{g}}$ is the Green's function of the effective medium calculated using the \texttt{almaBTE} code \cite{almaBTE2017}. From  $\bm{\mathsf{T}}_i$ the elastic scattering rates for each scatterer,  $\tau_{dis}^{-1}(i)$,  can be conveniently evaluated with the aid of the optical theorem, following the procedure used for point defects \cite{Mingo2010}. Finally, $\tau_{dis}^{-1}$ is obtained by a concentration average of $\tau_{dis}^{-1}(i)$ according to Matthiessen's rule.
More details on the symmetrization procedure and the Green's function formalism used to calculate the elastic scattering rates can be found in the Supplementary Materials.

\begin{figure}[bt]
\centering
\includegraphics[width=0.49\textwidth]{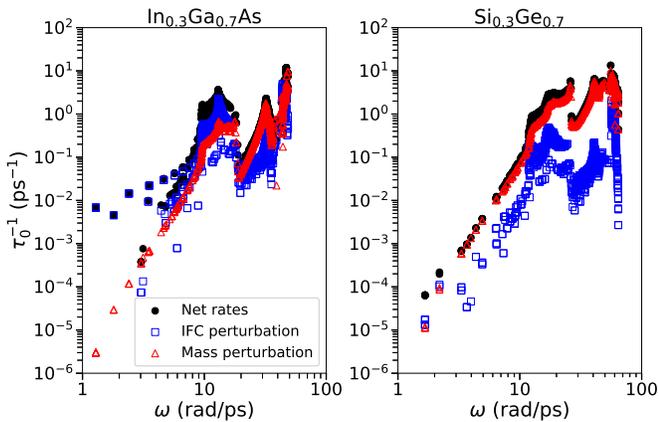} % The . refers to the current directory of the main tex file
\caption{Phonon-alloy elastic scattering rates calculated for \Chem{In_{0.3}Ga_{0.7}As} (left) and \Chem{Si_{0.3}Ge_{0.7}} (right) considering different types of perturbations: mass + force-constant disorder (black-filled circles), force-constant-disorder (empty blue squares) and mass disorder (empty red triangles). }
\label{fig:ingaas_vs_sige}
\end{figure}

We built several SQSs corresponding to different concentrations of \Chem{In_{1-x}Ga_xAs} random alloys (x = 0.25, 0.5, 0.7 and 0.85)  mostly in the \Chem{Ga}-rich region, where the $\kappa_{\ell}$ predicted by the VCA differs the most from the experiments.  In particular, for \Chem{In_{0.3}Ga_{0.7}As} we considered two different SQSs in order to evaluate to what extent the results are affected by the choice of the model system. As displayed in Fig. \ref{fig:kappal}, the two different SQSs  lead to values of $\kappa_{\ell}$ which differ by less than \SI{0.4}{\watt \per \meter \per \kelvin}.
 Figure \ref{fig:kappal} also shows that employing 128-atom SQSs is already sufficient to obtain a value of $\kappa_{\ell}$ in a very good  agreement with the experimental data. 

One can notice that, not only does the value of  $\kappa_{\ell}$ calculated through the VCA  quantitatively disagree with the experimental values, and with the results of our calculations, but there is also a qualitative difference in the behavior of $\kappa_{\ell}$ with the alloy composition. From Fig. \ref{fig:kappal}, one can see that, as \Chem{InAs} is characterized by a noticeably lower $\kappa_{\ell}$ than \Chem{GaAs}, the VCA predicts the minimum of $\kappa_{\ell}$ to be in the \Chem{In}-rich region, around $x_\mathrm{Ga} =0.35$. On the other hand, the experimental measurements and our calculations, suggest that the minimum is located in the region where the \Chem{In} and \Chem{Ga} concentration in the alloy is around 50\%. This behavior confirms that the complex atomic configuration characteristic of a random alloy affects the harmonic IFCs in a non-trivial way which  cannot be simply approximated through  a linear interpolation, as done in the VCA.

Finally, to emphasize the similarities and differences between our approach and the VCA, we also considered \Chem{Si_{1-x}Ge_x} by building a single 128-atom SQS of \Chem{Si_{0.3}Ge_{0.7}}. For this compound, as already mentioned, differences in $\kappa_{\ell}$ between our method and the VCA amount to less than the 10 \%. The analogous of Fig. \ref{fig:kappal} for \Chem{Si_{1-x}Ge_x} is reported in the Supplementary Materials.

The different predictive power that the VCA shows between  \Chem{In_{1-x}Ga_xAs} and \Chem{Si_{1-x}Ge_x} can be understood considering Fig. \ref{fig:ingaas_vs_sige}.
 The picture compares the values of $\tau_{dis}^{-1}$, calculated with our method, in \Chem{In_{0.3}Ga_{0.7}As} and \Chem{Si_{0.3}Ge_{0.7}}. The scattering rates are obtained considering the different terms in the expression of  $\langle \bm{\mathsf{V}}_i \rangle = \langle \bm{\mathsf{M}}_i + \bm{\mathsf{K}}_i \rangle$. 
We notice that for both compounds, for frequencies above  \SI{16}{\radian \per \pico \second} $\approx$ \SI{2.5}{\tera \hertz}, the effect of the IFC disorder perturbation on the scattering rates is small and the mass perturbation is the dominant term. For \Chem{Si_{1-x}Ge_x}, this is true also below   \SI{16}{\radian \per \pico \second} except for a very small region around  \SI{1}{\radian \per \pico \second}, where the rates are  close to zero anyway. On the other hand, \Chem{In_{1-x}Ga_xAs} is characterized by a  strong influence of the IFCs disorder perturbation on $\tau_{dis}^{-1}$  even below \SI{16}{\radian \per \pico \second}, corresponding to the region of the acoustic phonons frequencies. The effect of this phenomenon on $\kappa_{\ell}$ is particularly relevant since acoustic phonons  are the main heat carriers. This fact emphasizes the inability of a simple mass perturbation to correctly describe alloy disorder in \Chem{In_{1-x}Ga_xAs} and ultimately explains the failure of the VCA approach for this material. On the other hand, if the IFC disorder in the alloy is small, the mass disorder will dominate the scattering processes and the VCA gives reliable results, as is the case with \Chem{Si_{1-x}Ge_x}.  
In the Supplementary Materials we show that, in \Chem{In_{1-x}Ga_xAs}, first-order perturbation theory and  the  expression of the T-matrix reported in Equation~\eqref{eq:t-matrix}  give slightly different values of the elastic rates  for frequencies around \SI{10}{\radian \per \pico \second}, leading to a difference in $\kappa_{\ell}$ of less than \SI{0.3}{\watt \per \meter \per \kelvin}. The mass disorder perturbation is, on the other hand, always small enough that  the truncation of the Born series to its first term reproduces a value of $\tau_{dis}^{-1}$ with the same accuracy as the complete series expansion.  

\begin{figure}[bt]
\centering
\includegraphics[width=0.49\textwidth]{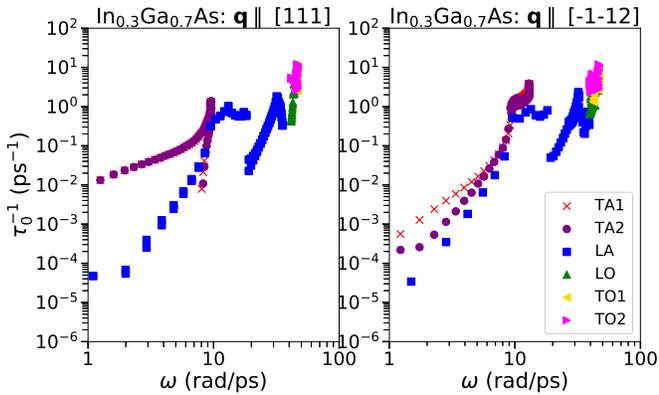} 
\caption{Phonon-alloy elastic scattering rates calculated for \Chem{In_{0.3}Ga_{0.7}As} for phonons with wavevector parallel (left) or perpendicular (right) to the \Chem{Ga/In-As} NN dipole. TA, LA, TO, LO indicate branches where, respectively, transverse acoustic, longitudinal acoustic, transverse optic and longitudinal optic phonons belong to. }
\label{fig:rates_vs_branch}
\end{figure}

The source of the exceptionally strong phonon-alloy scattering found for \Chem{In_{1-x}Ga_xAs} below \SI{16}{\radian \per \pico \second} partially originates from the polar nature of the system, as shown in Fig. \ref{fig:rates_vs_branch}. 
Here the elastic scattering rates are represented with different colors and symbols according to the phonon branch involved in the scattering process. On the left-hand side we depict incident phonons with wavevector parallel to the [111] direction, along which lies the  NN cation-anion  dipole of the zincblende structure, and on the right-hand side phonons with wavevectors along the [-1,-1,2] direction, which lies perpendicular to the [111]. When the phonon wavevector is parallel to the dipole, transverse acoustic phonons (TA) are scattered much more strongly than longitudinal ones (LA), giving rise to the intense scattering rates shown in Fig. \ref{fig:ingaas_vs_sige} for frequencies below \SI{16}{\radian \per \pico \second}. The dipole is however less perturbed by TA phonons traveling in a perpendicular direction and the difference between the rates of TA and LA phonons is significantly smaller. 

In conclusion, we have presented a general first-principles method for calculating $\kappa_{\ell}$ of random semiconductor  alloys which includes the contribution of IFC disorder in the description of the alloy scattering events. We have shown that  this inclusion  is necessary in compounds such as \Chem{In_{1-x}Ga_xAs}, characterized by polar bonds and strong fluctuations of the atomic-scale structure from the one assumed in the VCA.  On the other hand, in other compounds, such as \Chem{Si_{1-x}Ge_x},  where mass disorder is the leading factor, the VCA is a reasonable approximation. Overall, our new method can be implemented with a reasonable amount of computational resources, is compatible with the existing phonon BTE formalism and paves the way for the first-principles determination of $\kappa_{\ell}$ in general III-V and II-VI random semiconductor alloys.

We acknowledge support from the European Union's Horizon 2020 Research and Innovation Programme, grant number 645776 (ALMA). We thank the Vienna Scientific Cluster for providing the computational facilities (project number 70958: ALMA).

\end{document}